\documentstyle[12pt,epsf]{article}
\def\fnote#1#2{\begingroup\def\thefootnote{#1}\footnote{#2}\addtocounter
{footnote}{-1}\endgroup}

\def\inbar{\vrule height1.5ex width.4pt depth0pt}
\def\IB{\relax{\rm I\kern-.18em B}}
\def\IC{\relax\,\hbox{$\inbar\kern-.3em{\rm C}$}}
\def\ID{\relax{\rm I\kern-.18em D}}
\def\IE{\relax{\rm I\kern-.18em E}}
\def\IF{\relax{\rm I\kern-.18em F}}
\def\IG{\relax\,\hbox{$\inbar\kern-.3em{\rm G}$}}
\def\IH{\relax{\rm I\kern-.18em H}}
\def\II{\relax{\rm I\kern-.18em I}}
\def\IK{\relax{\rm I\kern-.18em K}}
\def\IL{\relax{\rm I\kern-.18em L}}
\def\IM{\relax{\rm I\kern-.18em M}}
\def\IN{\relax{\rm I\kern-.18em N}}
\def\IO{\relax\,\hbox{$\inbar\kern-.3em{\rm O}$}}
\def\IP{\relax{\rm I\kern-.18em P}}
\def\IQ{\relax\,\hbox{$\inbar\kern-.3em{\rm Q}$}}
\def\IR{\relax{\rm I\kern-.18em R}}
\def\ZZ{\relax{\sf Z\kern-.4em Z}}
\def\fnote#1#2{\begingroup\def\thefootnote{#1}\footnote{#2}\addtocounter
{footnote}{-1}\endgroup}
\def\beq{\begin{equation}}
\def\eeq{\end{equation}}
\def\bea{\begin{eqnarray}}
\def\eea{\end{eqnarray}}
\def\lleq#1{\label{#1}\eeq}

\let\nn=\nonumber

\def\notin{\ \hbox{{$\in$}\kern-.51em\hbox{/}}}
\def\a{\alpha}        
 \def\G{\Gamma}    \def\l{\lambda}
  \def\om{\omega}  \def\Om{\Omega} 
   
  \def\cC{{\cal C}}

\def\lra{\longrightarrow}

\def\hbar{\bar h}

\def\Inulls{{\bf I_0^*}} \def\II{{\bf II}} \def\III{{\bf III}}
\def\I{{\bf I}} \def\IV{{\bf IV}}
\begin{document}
\baselineskip=16pt
\parskip=.1truein
\parindent=0pt

\hfill {hep--th/9512138}
\vskip -.1truein
\hfill {NSF--ITP--95--158}
\vskip -.1truein
\hfill {BONN--TH--95--22}

\vskip .8truein

\centerline{\large {\bf Heterotic Gauge Structure of Type II
       K3 Fibrations}}

\vskip .5truein
\centerline{\sc Bruce Hunt$^1$
                 \fnote{\diamond}{Email: hunt@mathematik.uni-kl.de }
                and
                Rolf Schimmrigk$^{2,3}$
                   \fnote{\dagger}{Email: netah@avzw02.physik.uni-bonn.de}
            }

\vskip .3truein
\centerline{\it $^1$Fachbereich Mathematik, Universit\"at}
\vskip .01truein
\centerline{\it Postfach 3049, 67653 Kaiserslautern}
\vskip .1truein
\centerline{\it $^2$Institute for Theoretical Physics}
\vskip .01truein
\centerline{\it University of California, Santa Barbara, CA 93106}
\vskip .1truein
\centerline{\it $^3$Physikalisches Institut, Universit\"at Bonn}
\vskip .01truein
\centerline{\it Nussallee 12, 53115 Bonn}

\vskip 1.1truein
\centerline{\bf Abstract}

\noindent
We show that certain classes of K3 fibered Calabi-Yau manifolds
derive from orbifolds of global products of K3 surfaces and particular
types of curves. This observation explains why the gauge groups of
the heterotic duals are determined by the structure of a single
K3 surface and provides the dual heterotic picture of conifold
transitions between K3 fibrations.
Abstracting our construction from the special case of K3 hypersurfaces
to general K3 manifolds with an appropriate automorphism, we show how
to construct Calabi-Yau threefold duals for heterotic theories with
arbitrary gauge groups. This generalization reveals that the previous
limit on the Euler number of Calabi-Yau manifolds is an artifact of the
restriction to the framework of hypersurfaces.

\renewcommand\thepage{}
\newpage

\baselineskip=14pt
\parskip=.1truein
\parindent=20pt
\pagenumbering{arabic}

\noindent
{\bf 1. Introduction}

\noindent
It has been recognized recently that the agreement found in
\cite{kv95}\cite{fhsv95}\cite{hettype2} between the perturbative
structure of the
prepotentials for a number of heterotic string K3$\times $T$^2$
vacua and certain type II Calabi--Yau backgrounds can be traced back
to the K3-fiber structure of the models considered
\cite{klm95}\cite{vw95}\cite{fibs}.
Evidence based on the analysis of the weak coupling form of the
prepotential however is not convincing in the light of recent
discussions \cite{bkk95a,cggk95,acjm95} from which one learns that
moduli spaces of different Calabi--Yau manifolds intersect in
certain submanifolds. Thus weak coupling
arguments would appear insufficient to identify
heterotic duals\fnote{1}{A detailed investigation of this problem
will appear in ref. \cite{bkk95b}.}. This makes it particularly
important to develop different tools for identifying heterotic and
type II vacua which do not rely on a comparison of the perturbative
couplings.

In the present paper we wish to describe a way to identify heterotic
duals directly from the structure of the K3 fibrations and vice versa.
Instead of analyzing the prepotentials we focus
on the detailed geometry of K3 fibered Calabi--Yau manifolds which
turns out to contain sufficient information to derive
the heterotic gauge structure. The basic observation is that the
manifolds which have been encountered so far in the
context of heterotic/type II duality can in fact be described
as orbifolds of product manifolds defined by a K3 surface and
an appropriately defined curve. This shows that the essential
information of the fibration is determined by a single K3 surface
and thus provides an explanation of the fact that the gauge groups
of the heterotic duals of K3 fibered Calabi-Yau spaces are determined
by the singularity structure of K3 manifolds\fnote{2}{The relation
in the context of N$=$4 theories between
  the ADE singularities of K3 and ADE gauge groups of the heterotic
  dual on the torus has been explicated in \cite{ew95a}. Using the adiabatic
  limit it was argued in \cite{vw95} that this relation carries over
  to the N$=$2 framework.}.
Our construction then identifies the heterotic gauge group
of these theories
as the invariant part of the Picard lattice of the K3 fiber
with respect to the group action which gives rise to the fibration.
We will also see that the combination of this result with the
conifold transitions between K3 fibered
Calabi-Yau manifolds introduced in \cite{ls95}, and the analysis of
the origin of the gauge group in D$=$6 theories presented in
\cite{ew95a}, provides complete control of the dual heterotic picture
of the conifold transition on the Calabi-Yau manifold.

We can then turn around this observation and start from abstractly
defined orbifolds in which the fibers are not described by some
weighted hypersurface, or complete intersection, as has been
assumed in the most general class of fibrations presented so far.
This will allow us to construct Calabi--Yau manifolds for arbitrary
gauge groups. It turns out that the theory is most easily understood in
terms of elliptic fibrations of K3 surfaces in which the generic fiber
is a torus.
There are finitely many singular fibers, which were
classified by Kodaira in the sixties. This classification is related to the
classification of the simple rational doublepoints, and through this to the
classification of the simple Lie algebras. From our present standpoint we
can see another reason for this coincidence  --
we will find K3 surfaces with
elliptic fibrations (elliptic K3's) which can be used to construct
Calabi-Yau threefolds with K3 fibrations, which correspond dually to
heterotic strings. Since the unbroken gauge group of the heterotic string
is $E_8\times E_8$, the broken gauge group is a subgroup of this. The
correspondence between the singular fibers of the K3 and the lattice of
the gauge group of the heterotic string then dictates that the
possible singular fibers must also mirror
the classification of the gauge group.

A further result we find along the way concerns the possible limits
on the Hodge numbers of Calabi--Yau manifolds, i.e. the number of
vector multiplets and hypermultiplets of the heterotic theory.
Applying our construction to other K3's than the weighted
hypersurfaces we find that the limit found in the context of
weighted hypersurfaces is
not in fact a characteristic of Calabi--Yau manifolds per se.
In \cite{cls90}  weighted hypersurfaces
$\IP_{(1,1,12,28,84)}[84]$ and its mirror\fnote{3}{It can be
shown using the mirror transform of \cite{ls90,ls95} that
these manifolds are indeed mirrors.}
$\IP_{(11,41,42,498,1162,1743)}[3486]$, have been constructed
which define the `boundaries' of the mirror plot of \cite{cls90}
with the largest absolute value of the Euler number, namely 960.
This value has proved to be rather robust over the
 years, turning up in different constructions, such as
various classes of exactly solvable string vacua based on minimal
N$=$2 \cite{adecon} and Kazama-Suzuki models \cite{kscon},
the class of all Landau--Ginzburg theories \cite{ks94},
their abelian orbifolds \cite{kss92}, abelian orbifolds with
torsion \cite{krsk94}, as well as toric constructions
\cite{vvb94,cdk95}.
The fact that the space $\IP_{(1,1,12,28,84)}[84]$
has the maximal Euler number in the class
of Calabi--Yau hypersurfaces can be traced back to the fact
that the smooth fiber of the K3 fibration has an automorphism
of order 42. Results of Nikulin \cite{vvn79} show that
this is not the highest possible value. There are in fact two higher
values for the order of a K3 surface and we will discuss both
examples in this paper. Applying our construction to them we will find
Calabi-Yau threefolds with Euler number smaller than --960.
This shows that the
structure of the mirror plot of \cite{cls90} is  an artifact of the
construction after all.

\vskip .2truein
\noindent
{\bf 2. A Class of K3 fibered Calabi-Yau Manifolds}

\noindent
Before describing the geometry of general K3 fibrations more abstractly we
explain the essential ingredients in the more familiar framework of weighted
complete intersection manifolds.
This will allow us to be fairly concrete. Thus consider the class of
fibrations
\beq
\IP_{(2k_1-1,2k_1-1,2k_2,2k_3,2k_4)}[2k]
{}~\sim ~\matrix{\IP_{(1,1)}\hfill \cr
        \IP_{((2k_1-1),(2k_1-1),k_2,k_3,k_4)}\cr}
\left[\matrix{2&0\cr (2k_1-1) & k\cr}\right],
\lleq{k3fibs}
with $k=(2k_1+k_2+k_3+k_4-1)$ and $2k/(2k_1-1) \in 2\IN$, described
in \cite{ls95}\fnote{4}{As emphasized in \cite{ls95}, the assumption
            $2k/(2k_1-1) \in 2\IN$ is merely a matter of convenience
        and dropping this condition merely complicates the discussion,
   as explicated with a concrete example. Recently a heterotic/type II
  dual pair based on such a more general space with $k_1=2$, $k_2=4$,
   $k_3=10$ and $k_4=7$  has been discussed in detail in
  \cite{ks95}.}.
The relations (\ref{k3fibs}) which can be established via Landau--Ginzburg
considerations utilizing fractional transformations \cite{ls90,ls95}, are
useful for the analysis of the Yukawa structure of heterotic/type II
theories.

In the following we will also be needing information about the
the structure of the second cohomology
group\fnote{5}{The necessary ingredients for the following remarks
     can be found for example in \cite{res}.}.
The K\"ahler sector of the theory receives contributions from
two different sources.
First there is the (1,1)--form which comes from the restriction
of the K\"ahler form of the ambient space. Next there is a fixed
$\ZZ_2$--curve
\beq
\ZZ_2:~C=\IP_{(k_2,k_3,k_4)}[k]=\{p(z_2,z_3,z_4)=0\},
\eeq
the resolution of which adds one further contribution to the K\"ahler
sector.
This curve lives in a weighted projective plane, whose resolution
introduces a further number $N_C$ of (1,1)--forms,
depending on the relative divisibility properties of the weights of
the curve as well as the type of its defining polynomial.
Finally there is  the $\ZZ_{(2k_1-1)}$ fixed point set. The precise
structure of this set of  singularities depends on the divisibility
properties of $(2k_1-1)$ relative to the rest of the weights. To be
concrete we will present our discussion assuming
gcd$(2k_1-1,2k_i)=1$, $i=2,3,4$ (other situations being described
by modifications which can be easily derived using the information
contained in \cite{res})\fnote{6}{In the example $\IP_{(3,3,8,20,34)}[68]$
       the $\ZZ_3$--singular set e.g. is just a curve $\IP_1$).}.
 In such a situation the singular set is
\beq
\ZZ_{2k_1-1}:~~ \IP_1[2k/(2k_1-1)] = \frac{2k}{2k_1-1}~{\rm pts},
\eeq
the resolution of which leads to an additional $(2k(k_1-1)/(2k_1-1))$
(1,1)--forms. Thus we find a total of
\beq
h^{(1,1)} = 1 + \frac{(k_1-1)2k}{2k_1-1} + 1 + N_C
\eeq
(1,1)--forms.

The generic fiber is an element in the K3 configuration
\beq
K_{\l} = \IP_{(2k_1-1,k_2,k_3,k_4)}[k] \ni
\{(1+\l^{2k/(2k_1-1)})z_1^{k/(2k_1-1)}+p(z_2,z_3,z_4)=0\},
\lleq{k3-hyp}
which can be chosen to describe a quasismooth surface everywhere on
the base $\IP_1$ except at the $2k/(2k_1-1)$ points $\l_i$ which solve
$(1+\l_i^{2k/(2k_1-1)})=0$. Over these points the fibers degenerate.
Important for the following will be detailed structure of these
degenerations.
One of essential features of the class of manifolds (\ref{k3fibs})
is that the structure of the fibers does not change as one moves in
the fiber, they are of constant modulus. They do change rather
drastically however when one hits
one of the $2k/(2k_1-1)$ base points $\l_i$ on $\IP_1$ at which
the fibers degenerate.
 At these points the coordinate $z_1$ is completely unrestricted
and the degenerate fibers are cones over the curve
$C$ embedded in the generic fiber $K$.

The discussion so far suffices to derive the Euler number of the
spaces (\ref{k3fibs}) purely in terms of the fiber structure,
a result which we will have use for later on. The necessary ingredients
of this computation are the Euler number of the base,
$\chi(\IP_1)=2$, the Euler number of generic fiber, $\chi(K3)=24$,
the number $(2k/(2k_1-1))$ of singular base points,
and the Euler number of the degenerate fiber. The structure of the
singular fibers depends crucially on whether $k_1$ is equal or larger
than unity.  If $k_1> 1$ there is the additional complication that
 each vertex of the cone over $C$ is
a $\ZZ_{2k_1-1}$--singular point on the Calabi-Yau manifold whose
resolution introduces $(k_1-1)$ spheres $\IP_1$. Thus the Euler number of
the degenerate fibers $F_{\rm deg}$ is given by
\beq
\chi(F_{\rm deg}) = \chi(C) + 2(k_1-1) +1
\eeq
and therefore the Euler number of the fibered threefold
\beq\label{chiformula}
\chi(M)
= (2-N_s)\cdot 24 + N_s (\chi(C) + 2k_1 -1).
\lleq{fib-euler}

The second crucial property of the  manifolds (\ref{k3fibs}) is that the
monodromy transformation {\bf m}, generated by
\beq
\ZZ_{k/(2k_1-1)} \ni {\bf {\rm m}}: ~~
(z_1,...,z_4)~~\lra ~~(\a z_1,z_2,z_3,z_4),
\lleq{mono-act}
is nilpotent of degree $k/(2k_1-1)$, i.e. {\bf m}$^{k/(2k_1-1)}=1$.

The structure of the fibrations (\ref{k3fibs}) explicated thus far allows
us to draw on some general results of birational geometry\fnote{7}{For a
  single nilpotent fiber, by definition, the monodromy satisfies
  ${\bf m}^s=1$, hence a ramified cover of the base, branched to order $s$ at
  the base point of the singular fiber, has trivial monodromy. For a global
  fiber space with nilpotent monodromy, the same holds for a
  suitably chosen ramified cover of the base.}
in order to get further insight into their structure. Namely, since the
monodromy is nilpotent and the modulus is constant, it follows
 that these manifolds can in fact be described (birationally) as
orbifolds of products\fnote{8}{On the ramified cover of the last footnote,
  the monodromy is trivial, and the modulus is constant; this means we have
  a product.} of the form $\cC_{k/(2k_1-1)} \times K$,
where $p: \cC_{k/(2k_1-1)} \lra \IP_1$ is the projection of a
$k/(2k_1-1)$--fold cover of the base space of the fibration.
In order to see this consider the action of the cyclic group
$\ZZ_{k/(2k_1-1)}$ on the product
\beq
\ZZ_{k/(2k_1-1)}:~\cC_{k/(2k_1-1)} \times K ~\lra
                    \cC_{k/(2k_1-1)}\times K
\eeq
defined by the projection $p$ on the first factor and the monodromy
action {\bf m} (\ref{mono-act}) on the second factor.
The action {\bf m} leaves invariant the curve
$C$ and therefore the orbifold
$\ZZ_{k/(2k_1-1)}\backslash \cC_{k/(2k_1-1)}\times K$
will have $2k/(2k_1-1)$ singular fibers which
are obtained by blowing up the curve $C$ in the fiber $K_{\l_i}$
($\l_i$ being any of the $2k/(2k_1-1)$ branch points of
 $\cC_{k/(2k_1-1)}$), resulting in a ruled surface\fnote{9}{A ruled
 surface is a fibration over a curve (here $C$) with
  fiber $\IP_1$.} $E_i$.  The surface $K_{\l_i}$ has as quotient
the weighted projective plane $\IP_{(k_2,k_3,k_4)}$ while each
$E_i$ descends to the orbit space (being fixed under the action of
${\ZZ}_{k/(2k_1-1)}$), and is in the branch locus there.
Thus on the resolved orbit
space each singular fiber over a $k/(2k_1-1))^{st}$ root of unity
consists of two components, a plane $\IP_{(k_2,k_3,k_4)}$
and a ruled surface $E_i$, the two intersecting in the curve
$C$. Because $C$ is just a hyperplane section of the original
fiber $K_{\l_i}$, it follows that the weighted projective plane
can be blown down to a point. In this process the intersection
curve $C$ is blown down to a point as well and the surface
$E_i$ becomes a cone over the curve $C$. This is precisely the
structure we have previously found for the manifolds (\ref{k3fibs})
and thus we have uncovered that the essential structure of the weighted
hypersurfaces (\ref{k3fibs}) is that of orbifolds of a global product
involving K3 surfaces.

This may be described more explicitly as follows. Let
$\pi:{\cal C}_{k/(2k_1-1)}\lra \IP_1$ denote the projection. We
may set $z_1=y_1^{1/2}$ here because all weights are divisible by 2.
Then define (with $\ell=k/(2k_1-1)$)
\begin{eqnarray}
\phi:\ZZ_{\ell}\backslash \cC\times K &
  \hookrightarrow & \IP_{(2k_1-1, 2k_1-1, 2k_2, 2k_3, 2k_4)} \nn \\
(x_0,(x_1,z_2\ldots, z_4))~{\rm mod}~\ZZ_{\ell} & \mapsto &
  \left({\pi(x_0)\over (1+\pi(x_0)^{2\ell})^{1/2\ell}}x_1^{1/2},
   {1\over (1+\pi(x_0)^{2\ell})^{1/2\ell}}x_1^{1/2}, z_2,z_3,z_4\right). \nn \\
\end{eqnarray}
Here the map is defined for $\pi(x_0)^{2\ell}\neq -1$;
we then take the closure in the projective space.
Then clearly $\phi(x_0,K)$ is the hyperplane section
$z_1=\pi(x_0)z_2$ of
\beq
\{f(z_1,\ldots,z_5)=
  z_1^{2k/(2k_1-1)}+z_2^{2k/(2k_1-1)}+p(z_2,z_3,z_4)=0\}
  \subset\IP_{(2k_1-1, 2k_1-1, 2k_2, 2k_3, 2k_4)}.
\eeq
It is well defined on the quotient because only $\pi(x_0)$ occurs, and for
$\pi(x_0)^{2\ell}=-1$ the image is the cone
\beq
 C=\{p(z_2,z_3,z_4)=0\}\subset
\{z_1=\pi(x_0)z_2\}\subset \IP_{(2k_1-1, 2k_1-1, 2k_2, 2k_3, 2k_4)}.
\eeq
In this way the birational map described above is immediately performed
and in particular one can determine whether the vertex of the cone
$H_{\pi(x_0)}$ for $\pi(x_0)^{2m_1}=-1$, namely the point
$(\pi(x_0),1,0,0,0)$ is
a singular point of the threefold. Looking at the equation, it is clear
that the vertex is a quasi-smooth point of the threefold, and in
particular, if $k_1=1$, it is a smooth point, while if $k_1>1$, it is a
singular point of the ambient projective space and must be resolved.

Combining the results of the previous paragraphs
shows that the cohomology of the generic fiber
of the fibration
\beq
\matrix{\IP_{(2k_1-1,k_2,k_3,k_4)}[k] &\lra
                           & \IP_{(2k_1-1,2k_1-1,2k_2,2k_3,2k_4)}[2k]\cr
                           &          &      \cr
                           &          &\downarrow \cr
                           &          &      \cr
                           &          &\IP_1 \cr}
\lleq{lsfibs}
receives contributions from both, the resolution of the orbifold
singularities as well as those forms spanned by the polynomial
ring. More precisely, the subgroup
H$^{(1,1)}(K) \subset $H$^2(K)$, which is 20-dimensional
for a K3 manifold, is spanned by the K\"ahler form of the
ambient space of the fiber configuration $\IP_{(2k_1-1,k_2,k_3,k_4)}[k]$,
the contribution $N_C$ of the resolution of the singular points lying
on the curve $C$, and the complex deformations.
The group  H$^2(K)$ is endowed with a natural
inner product given by $<\om, \eta>=\int \om \wedge \eta$ for
$\om, \eta \in$ H$^2(K)$, the signature of which is
(3,19),  with 3--dimensional positive definite subspace.

We now see that the structure of the second cohomology group of any
of the spaces of type (\ref{k3fibs}) is determined by a single
K3 hypersurface and the action of the automorphism. We thus have reduced
the problem of deriving the heterotic gauge structure to the problem
of deriving it from the K3. This is easily achieved by considering the
invariant part of the Picard lattice with respect to the action which
defines the fibrations. We will illustrate this with some examples in
the next Section.

\vskip .2truein
\noindent
{\bf 3. Hypersurface Examples}

\noindent
{\bf Example I:} Consider the manifold
$\IP_{(1,1,12,28,42)}[84]$ which has been playing a
somewhat distinguished role over the last five years because it,
together with its mirror, features the to date largest known
absolute value of
Euler number, defining the extreme boundary of the mirror plot
of \cite{cls90}.  The candidate heterotic dual has been discussed in
\cite{kv95,afiq95,hm95,fibs}.
This manifold has the $\ZZ_2$--singular curve
$C=\IP_{(6,14,21)}[42]$,
on top of which we have the orbifold ponts
$\ZZ_4: \IP_{(3,7)}[21]=1~pt$, $\ZZ_6: \IP_{(2,7)}[14]=1~pt$,
and $\ZZ_{14}: \IP_{(2,3)}[6]=1~pt$.
The resolution of these points introduces one, two, and nine (1,1)--forms
respectively and therefore we find, including the one (1,1)--form coming
down from the ambient form, that the smooth resolved space has a total of
$h^{(1,1)} = 11$ (1,1)--forms. Using $c_3=-164978h^3$
one finds for the Euler number $\chi = -960$.

Considering the K3 fibration of this space then shows that the second
cohomology of the generic fiber
\beq
\IP_{(1,6,14,21)}[42] \ni
K_{\l} = \{(1+\l^{84})y^{42} + z_3^7 +z_4^3+z_5^2=0\}
\lleq{oldmaxk3}
decomposes as $20=1+9+10$, the first form
coming from the ambient K\"ahler form of $\IP_{(1,6,14,21)}$,
the nine forms decomposing into
$9=1+2+6$ resolution modes, and finally the 10 monomials
\beq
z_1^{28-6n}z_2^n z_3,~0\leq n\leq 4;~~~~~
z_1^{42-6m}z_2^m,~1\leq m\leq 5.
\lleq{k3ring}
We therefore see that the manifold
$\IP_{(1,1,12,28,42)}[84]$ arises by choosing
$K=\IP_{(1,6,14,21)}[42]$ and
considering the action of $\ZZ_{42}$ on the product
$\cC_{42}\times K$.
The quotient $\ZZ_{42}\backslash\cC_{42}\times K$
has 84 singular fibers obtained by first blowing up the curve $C$ in
the fiber $K_{\l_i}$ and then blowing down the projective plane to a
point. The resulting degenerate fibers are cones over the curve
$C$ with $\chi_C=11$ and plugging the values of $N_C=84$ and $\chi_C$
into our fibration formula (\ref{fib-euler}) reproduces the known result.

Thus we should look for the invariant part of the Picard lattice of the
K3 hypersurface $K=K_0$ (\ref{oldmaxk3}) with respect to the group
$\ZZ_{42}$.  It is generated by the K\"ahler form descending
from the ambient space and the 9 modes coming from the resolution.
Since the remaining part of H$^2$(K3) transforms nontrivially, as can
easily be inferred from the transformation behavior of the
monomials (\ref{k3ring}),
the invariant sublattice of the K3 lattice $\G^{(3,19)}$
is determined by the intersection form of the K3
$\G^{(1,9)} \subset \G^{(3,19)}$.
This sublattice is a selfdual Lorentzian even lattice and decomposes
into $\G^{(1,1)}\oplus \G^{(0,8)}$ in which the second term
denotes the root lattice of the E$_8$.

The geometry of this situation is encoded by the resolution of the
curve $C$. As described above resolving the orbifold singularities
of the ambient space leads to a contribution of $N_C=9$, coming
from the resolution of the three singular points sitting on top of
the curve. The resolution of each of these points introduces the
Hirzebruch-Jung trees described by the diagrams $A_1$, $A_2$ and
$A_6$ respectively, and the curve $C$ glues these trees together,
resulting in the graph
\[\unitlength .5cm
\begin{picture}(8,7)
\put(0,1){\line(1,0){7.5}}
\put(.5,3){\line(1,0){2}}
\put(1.5,4){\line(1,0){2}}
\put(2.5,5){\line(1,0){4}}
\put(5.5,4){\line(1,0){2}}
\put(1,3.5){\line(0,-1){3}}
\put(2,4.5){\line(0,-1){2}}
\put(3,5.5){\line(0,-1){2}}
\put(4.5,5.5){\line(0,-1){2}}
\put(6,5.5){\line(0,-1){2}}
\end{picture}
\]
whose dual describes precisely the lattice  $E_8 \oplus U$,
where $U$ denotes the hyperbolic plane.
Thus we see that the heterotic dual should be determined by higgsing
the first E$_8$ completely while retaining the second E$_8$.
We also see that we should not fix the radii of the torus at some
particular symmetric point but instead embed the full gauge bundle
structure into the E$_8$.
Doing precisely as instructed by the manifold we recover the heterotic
model of \cite{kv95}.

\noindent
{\bf Example II:} Consider the manifold $\IP_{(1,1,2,4,4)}[12]$
whose Hodge numbers were found \cite{kv95} to match that of a particular
heterotic model. The detailed understanding of this space
is of particular interest because it is known to be connected via
a conifold transition to a codimension two
Calabi--Yau manifold \cite{ls95}.
We therefore wish to see whether we in fact can derive the
heterotic theory from this Calabi--Yau manifold.
For this we have to determine the invariant sector of the Picard
lattice of H$^2(K3)$ under the orbifolding. In order to do so we
only need to observe that the orbifold singularities of the curve
$\IP_{(1,2,2)}[6]$ are three $\ZZ_2$--points whose resolution leads to
a total of 3 (1,1)--forms which, together with the K\"ahler
form of the ambient space, determines the sublattice
$\G^{(1,3)} \subset \G^{(3,19)} = {\rm H}^2({\rm K3},\ZZ)$.
Taking into account the divisor coming from the curve $C$ we find
the resolution diagram
\[\unitlength .5cm
\begin{picture}(6,4)
\put(2,3){\line(0,-1){3}}
\put(1.5,2){\line(1,0){1}}
\put(1.5,1){\line(1,0){1}}
\put(1,.5){\line(1,0){4.5}}
\end{picture}
\]
with $\chi(C)=3$.
We see from this that the intersection matrix is precisely given by
the Cartan matrix of the group SO(8). We also see that
in the heterotic dual we need to take the torus at the SU(3) point
in the moduli space and break this SU(3) by embedding the K3
gauge bundle structure groups appropriately. In this way we recover
the heterotic construction of \cite{kv95}.

\vskip .2truein
\noindent
{\bf 4. The General Construction}

%\begin{figure}
% \epsfbox{cone1.eps}
%\caption{
%The left hand arrow is the blow up along the fixed curve $C$. The
%  right hand arrow is the blow down of the original quotient of the fiber
%  $K$.}
%\end{figure}

\noindent
In this Section we abstract from the framework of weighted spaces
and describe our construction for general K3 fibers which do not
necessarily admit a description as weighted hypersurfaces.
For this we make the following assumptions. We are given a smooth K3
surface $K$ with an automorphism group $\ZZ_k$, such that the fixed
point set of the group is a curve on $K$, that is, there are no isolated
fixed points. Then we use the curve ${\cal C}_k$ of Section 2, the
$k$-fold cover of $\IP_1$ branched at the $N$ roots of unity on
$\IP_1$, where $k$ divides $N$. A somewhat involved computation
 then shows that the quotient ${\cal C}_k\times K$ by $\ZZ_k$ will
be (birationally) Calabi-Yau if and only if
\beq N=2k.\eeq
This is the value we found above, and it is true more generally. Now the
quotient space $\ZZ_k\backslash {\cal C}_k\times K$ will be singular at
the curve $C \subset K$ which is the fixed point curve of the action on
$K$. This must be blown up, and the exceptional divisor is a ruled surface,
i.e., a $\IP_1$-bundle over $C$. After this blow up, the fiber over one
of the roots of unity consist in two components: the quotient of the
original fiber (that is a copy of $K$), and this ruled surface. This is
depicted in the middle picture of Figure 1. After this blow up, however,
the component which is the quotient of $K$ becomes {\it exceptional}, and
can be blown down. This is depicted by the second arrow of Figure 1.

\vspace*{1cm}
\centerline{\epsfbox{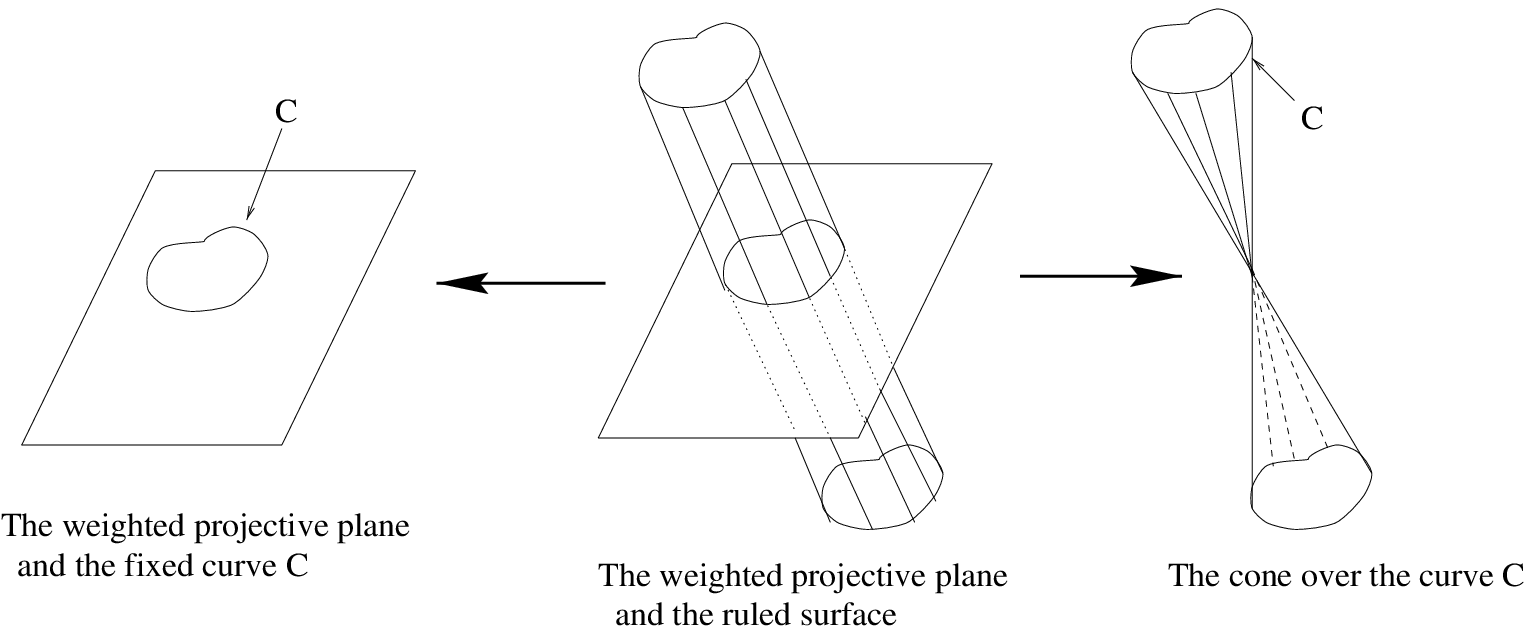}}
\noindent
{\bf Figure 1:}{\it ~The left hand arrow is the blow up along the fixed
        curve $C$. The right hand arrow is the blow down of the original
       quotient of the fiber $K$. }

The result is a cone over the curve $C$. Of course, the curve $C$ need not
be irreducible, and if there are several components, the resulting ruled
surface consists of several components. One can still
blow up the threefold along the curve $C$, and the result is a ruled
surface over the (reducible) curve $C$. This is depicted in Figure 2.

%\begin{figure}
%\epsfbox{cone2.eps}
%\caption{Here the fixed curve $C$ is reducible. In this case the situation
%  is the same, except that the ruled surface has several components now. }
%\end{figure}

\centerline{\epsfbox{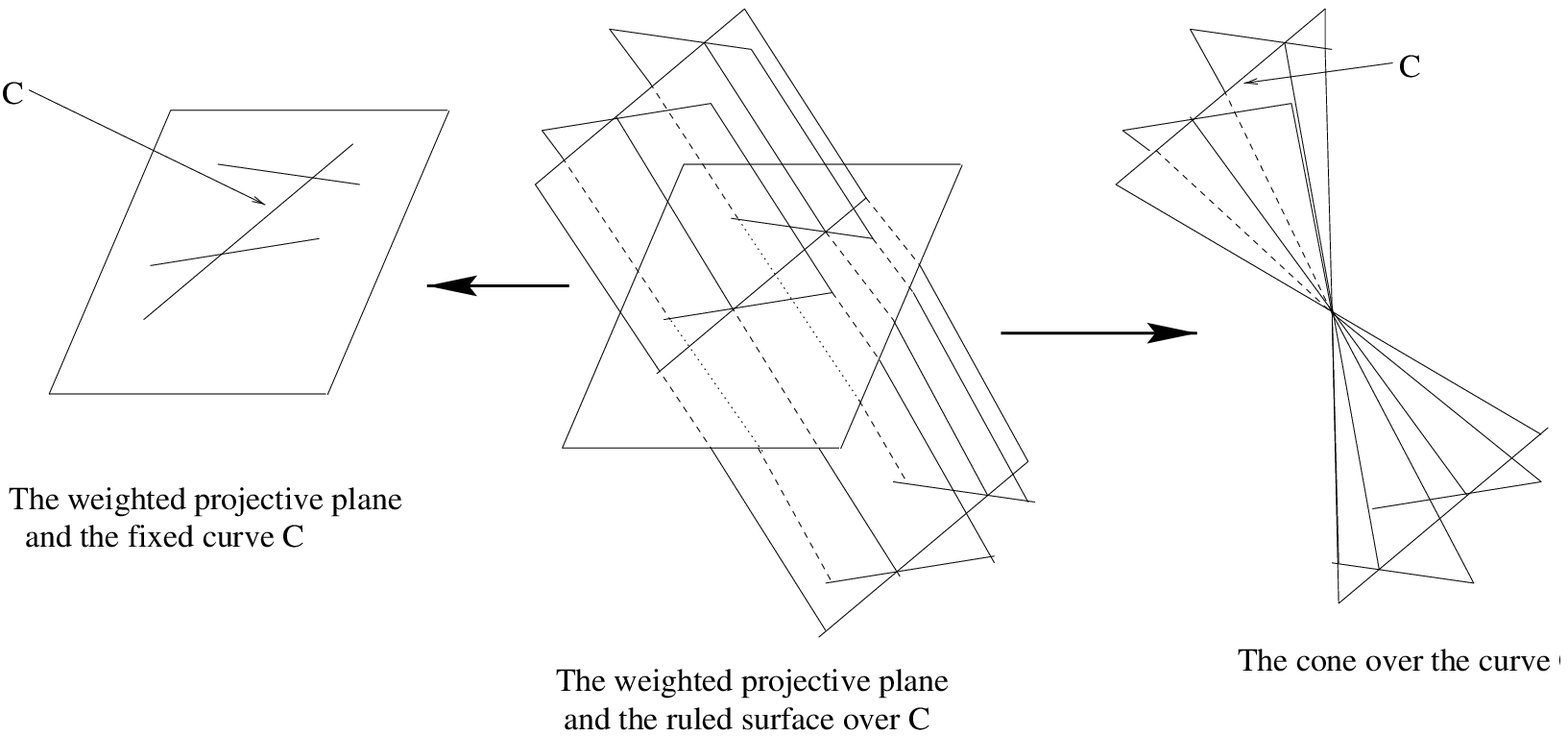}}
\noindent
{\bf Figure 2:}{\it ~Here the fixed curve $C$ is reducible. In this case
   the situation is the same, except that the ruled surface has several
   components now. }

After this blow down, the vertex of the cone may or may not be a singular
point of the threefold. We saw above that for the weighted hypersurfaces
with $k_1>1$, this was indeed a singular point and needed to be
resolved. The reason is that in those cases the ambient projective space
is singular at the vertex.
That is also the reason for the occurance of the degree $k_1$ in
the $\chi$-formula. For the elliptic surfaces considered below, this
problem does not occur.

Finally we mention that this construction, starting from just a smooth K3
with an automorphism, is applicable to {\it any} such surface, and in
particular, can be applied to elliptic K3 surfaces (see below) as well as
to weighted complete intersection K3's. This is especially important,
as the latter appear in the weighted conifold transitions
described in \cite{ls95}, generalizing
the splitting conifold construction of \cite{cdls88}.
Combining the results of \cite{ls95} with what we have learnt
so far allows us to gain a complete understanding of the dual heterotic
gauge structure of the conifold transition. More precisely we need to
collect the following ingredients: 1) The fibered Calabi-Yau threefold is
completely determined by a single K3 surface. 2) The conifold transition
connects a fibration with another fibration. For general conifold
transitions this will not be the case, but as shown in \cite{ls95}
there exist conifold transitions for which this holds. It was furthermore
shown in \cite{ls95} that such transitions proceed via a degeneration
of the fibers. 3) The singularity structure of the K3 surface determines
the dual heterotic gauge group \cite{ew95a}. Combining these facts we
see that in conifold transitions between K3-fibered threefolds the resolution
graph of the K3 surface changes because of the vanishing and appearance
of 2--cycles when the K3 fibers go through the degenerate configuration.
Since it is this graph which determines the Dynkin diagram we thus
gain an understanding of the heterotic dual of the transition.

\vskip .2truein

\noindent
{\bf 5. Automorphisms of K3 Surfaces}

Our discussion of the previous Sections shows that we need
 to understand the automorphisms of K3 surfaces, in particular when
group actions by some $\ZZ_k$ exist. The
first observation is that if $\Omega_K$ denotes the non-vanishing
holomorphic two form on $K$, then any automorphism $g$ acts via
$g^*\Omega_K = \alpha_K(g)\Omega_K$, where $\alpha_K(g)\in \IC^*$,
yielding an exact sequence
\beq
1 \lra G_K \lra Aut(K) \stackrel{\alpha_K}{\lra} \ZZ_k \lra 1,
\eeq
where $\ZZ_k$ is the cyclic group of $k^{th}$ roots of unity in
$\IC^*$ and $G_K$ is the kernel, i.e., the set (a group actually) of
automorphisms preserving the form $\Om_K$. This gives a representation
of $\ZZ_k$ in $T_K\otimes \IQ$, which is by results of Nikulin the
direct sum of irreducibles, of maximal rank $\phi(k)$, where $\phi$ denotes
the Euler function. More precisely, Nikulin's result \cite{vvn79}
is that all eigenvalues of $\ZZ_k$ acting on $T_K\otimes \IQ$ are
primitive $k^{th}$ roots of unity. Each irreducible component has the
maximal possible rank, namely $\phi(k)$.
Since $\phi(k)\leq rank(T_K)$, it follows that $k\leq
66$. Particularly interesting are the automorphisms that act trivially on
the Picard group. This group, denoted by $H_K$, is in fact
the $\ZZ_k$ as above (which shows the
sequence splits). As a consequence of this we learn that for an
element $g\in H_K$, the invariant lattice
under $g$ is precisely $S_K$, the Picard lattice.
It is clear from this that our main interest will be in elements of $H_K$,
so it is desirable to know more about what can possibly occur. A first
result in this direction was given by Kondo.
It was shown in \cite{sk92} that for unimodular $T_K$ $k$ must be
a divisor of any of the values in
$S=\{66, 44, 42, 36, 28, 12\}$. Furthermore, if $\phi(k)={\rm rk}(T_K)$
then $k$ takes precisely the values of $S$, and in these cases
  there exists a unique (up to isomorphism) K3 surface
  with given $k$.

For the examples indicated by Kondo's list $S$
 the invariant lattice $S_K$ and its
complement $T_K$ are as follows.
\[\begin{array}{c | c c} k & S_K & T_K \\ \hline
66 & U & U\oplus U \oplus E_8 \oplus E_8 \\
44 & U &  U\oplus U \oplus E_8 \oplus E_8 \\
42 & U\oplus E_8 &  U\oplus U \oplus E_8  \\
36 & U\oplus E_8 &  U\oplus U \oplus E_8  \\
28 & U\oplus E_8 &  U\oplus U \oplus E_8  \\
12 & U\oplus U \oplus E_8 & U \oplus U
\end{array}
\]

The situation for $G_K$ is just the opposite of the one
for $H_K$. The invariant
sublattice is this time $T_K$, and the action on $S_K$ was described for
abelian groups $G_K$ by Nikulin. The possible such $G_K$ which can occur
are the following:
 $$(\ZZ_2)^m, \ 0\leq m\leq 4;\ \ \ZZ_4;\ \ \ZZ_2\times \ZZ_4;\ \
 (\ZZ_4)^2;\ \ \ZZ_8;\ \ \ZZ_3;\ \ (\ZZ_3)^2;\ \ \ZZ_5;\ \
   \ZZ_7; \ \ \ZZ_6;\ \ \ZZ_2\times \ZZ_6.$$
Since the largest cyclic group occuring is $\ZZ_8$ it follows that
if  a K3 surface admits a cyclic automorphism
  of order $k\geq 9$, then this automorphism is in $H_K$.
\noindent Therefore, depending on our aims, it may be more useful to
consider $H_K$ or $G_K$. Note that by mirror symmetry (which for K3
surfaces is a theorem), there is another K3 surface $K^m$ for which $T_K$
and $S_K$ are exchanged. Consequently $H_K$ and $G_K$ are switched also.

We now apply these results to our construction.
Let $K$ be a K3 surface with an automorphism of order $k$ in $H_K$, and let
$S_K$ be the Picard lattice. Then with our construction above, we have: a
Calabi-Yau threefold $Y$ with a K3 fibration with $S_K$ is the invariant
part of the lattice. Hence the data $(K,\ZZ_k=H_K, S_K)$ determines a
Calabi-Yau, dual to a heterotic string with gauge group lattice which is
isomorphic to $S_K$. In other words, to identify the invariant
lattice, it is sufficient to determine the Picard lattice $S_K$ of
$K$. Note that, to get a K3 surface with a given $S_K$, it is sometimes
sufficient to give a combination of singular fibers and an elliptic K3
surface with those singular fibers.
Let us give a brief description of this class of K3 surfaces. An elliptic
curve can always be realized as a cubic in $\IP_2$. To get an elliptic
surface, one lets this cubic curve in the plane vary. This is described by
an affine equation of the type
\beq
y^2=x^3-g_2(t)x-g_3(t),
\eeq
where $g_i(t)$ is a section of a line bundle $L^{\otimes 2i}$ on some curve
$C$ (the base curve of the projection of the surface $S\lra C$). Here $x$
and $y$ are affine coordinates in a $\IP_2$, and the entire surface is
contained in a $\IP_2$-bundle over $C$. If $S$ is a K3 surface, then
necessarily $C=\IP_1$, and the sections $g_i(t)$ are just homogeneous
polynomials of degrees $2i\cdot deg(L)$, and for $S$ to be K3 again we need
$deg(L)=2$. The fiber over a point $t\in \IP^1$ will be singular
precisely when the discriminant of the
Weierstra\ss\ polynomial vanishes there,
$\Delta(t):=g_2(t)^3-27g_3(t)^2=0$. The type of singular fiber is
completely determined by the degrees of vanishing of $g_2,\ g_3$ and
$\Delta$ at the point, according to Table.
\begin{small}
\[\begin{array}{|l|c|c|c|c|c|c|c|c|c|c|} \hline
Fiber & \I_0 & \I_n, n>0 & \II & \III & \IV & \I_0^* & \I_n^*, n>0 & \II^*
& \III^* & \IV^* \\ \hline
\nu(g_2) & 0 & 0 & \geq 1 & 1 & \geq 2 &  2\  / >2\ / 2
& 2 & \geq 4 & 3 & \geq3 \\ \hline
\nu(g_3) & 0 & 0 & 1 & \geq 2 & 2 & >3\ /  3\ /  3
& 3 & 5 & \geq 5 & 4 \\ \hline \nu(\Delta) & 0 & n & 2 & 3 & 4
& 6 & n+6 & 10 & 9 & 8 \\ \hline
{\cal J} & \neq 0, 1, \infty & \hbox{order $n$ pole} & 0 & 1 & 0 &
1\ /  0 \ / \neq 0,1,\infty
& \hbox{order $n$ pole} &  0 & 1 & 0 \\ \hline
\end{array}
\]
\end{small}
\noindent
{\bf Table:}~{\it
In the last row we have listed the value of the ${\cal J}-$function, which
is defined by ${\cal J}=g_2^3/\Delta$. If ${\cal J}$ is constant, then the
modulus of the elliptic curve is constant in the family. The fibers are as
shown in Figure 3.}

Looking at the table, the following is clear. If we consider the dual graph
of each fiber type (with the exception of $\II$),
then we get an extended Dynkin diagram of one of the simple Lie algebras.
The correspondence is given as follows:
\[\begin{array}{ccc|c|c|c|c|c} \II & \III & \IV & \I_n & \I_n^* &
\IV^* & \III^* & \II^* \\ \hline
- & A_1 & A_2 & A_{n-1} & D_{n+4} & E_6 & E_7 & E_8
\end{array}
\]
(see \cite{sk92}). In this way it is often possible to see
what the lattice $S_K$ of such an elliptic surface is.
More precisely, if there is a unique
section, then the Picard lattice $S_K$ can be read off directly from the
singular fibers. This is the situation with all the examples of Kondo.

We now describe the three examples of Kondo which we shall use later. These
are all elliptic fibrations.

%\begin{figure}[h]\hspace*{1cm}
%\epsfbox{fibers.eps}
%\caption{In the first row are the fibers of types $\I_0$, $\II$, $\III$,
%  $\IV$ and $\I_n$, respectively. In the second row are the types
%  $\I_0^*$, $\II^*$, $\III^*$, $\IV^*$ and $\I_n^*$, respectively. The
%  fibers of the second row are the minimal resolutions of the quotients of
%  the fibers of the first row by the involution $z\mapsto -z$ of the
%  elliptic curve.}
%\end{figure}

\begin{figure}\hspace*{1cm}
\epsfbox{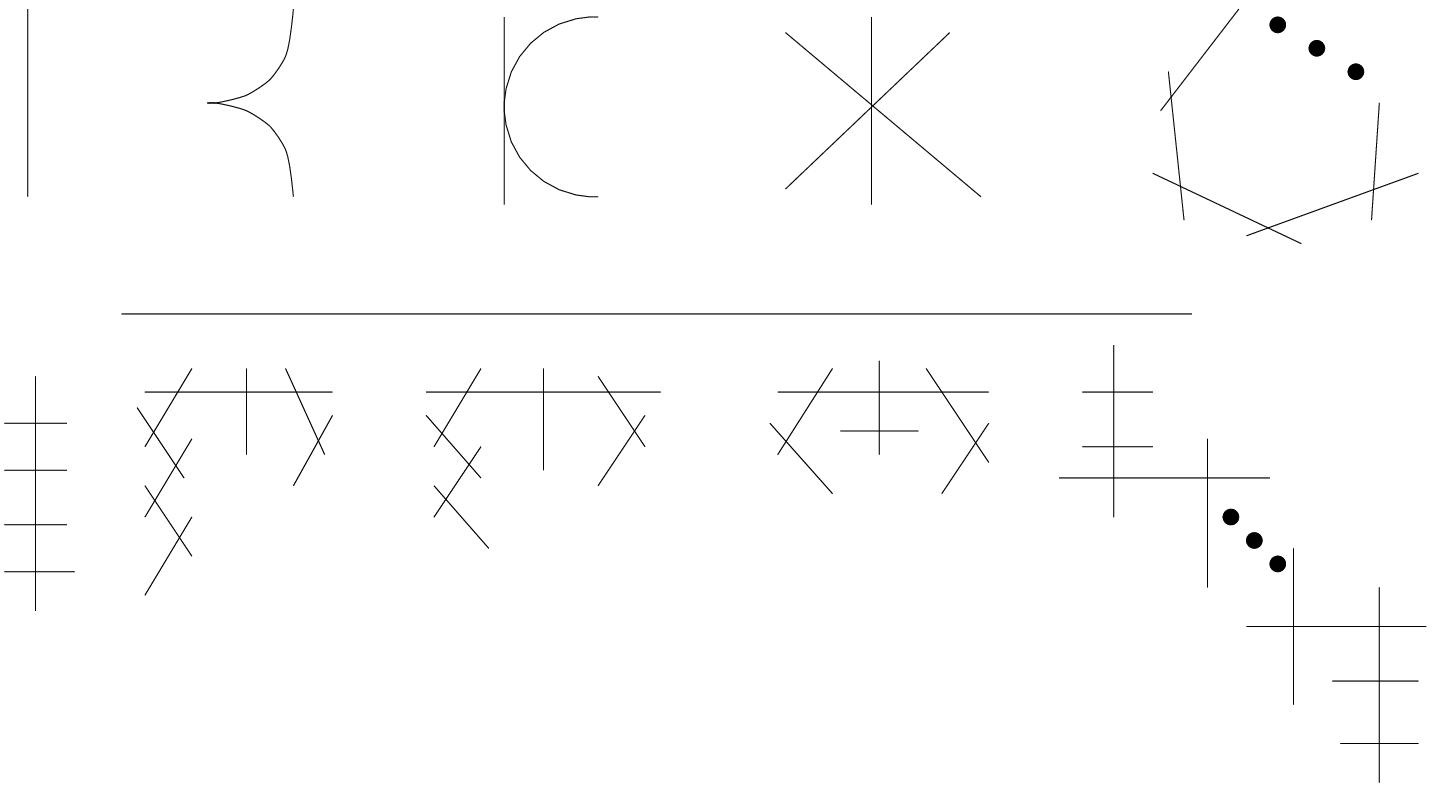}

\noindent
{\bf Figure 3:}~{\it
  In the first row are the fibers of types $\I_0$, $\II$, $\III$,
  $\IV$ and $\I_n$, respectively. In the second row are the types
  $\I_0^*$, $\II^*$, $\III^*$, $\IV^*$ and $\I_n^*$, respectively. The
  fibers of the second row are the minimal resolutions of the quotients of
  the fibers of the first row by the involution $z\mapsto -z$ of the
  elliptic curve.}
\end{figure}

\medskip
\noindent{$k=66$:}
There are two ways to construct this surface: Kondo describes it as an
elliptic K3, with 12 fibers of type \II\ at $t=0$ and at the $11^{th}$
roots of unity. The affine equation is
\beq
y^2=x^3+t\prod_1^{11}(t-\alpha_{11}^i),
\eeq
and the automorphism is given by
$(x,y,t) \mapsto (\alpha_{66}^2 x, \alpha_{66}^3 y, \alpha_{66}^6 t).$
Alternatively, one can consider the (non-Gorenstein) weighted hypersurface
\beq
\{x^2+y^3 +z^{11} + w^{66} =0\} \in \IP_{(1,6,22,33)}[66],
\eeq
which, upon resolution, yields a smooth K3 \cite{id95}.
Here the automorphism is given by
$(x,y,z,w) \mapsto (x,y,z,\alpha_{66} w)$.  From this second
description we see that the fixed point set is the (total transform of the)
curve $\{x^2+y^3+z^{11} =0 \}\subset \IP_{(6,22,33)}$.
In the above description that curve is given by at most the fibers
$\pi^{-1}(0)$ and $\pi^{-1}(\infty)$ as 0 and $\infty$ are the only
fixed points of $t$, together with the zero section, the locus
(in $\IP_1\times \IP_2$) given by setting $x=0$ and $y=0$.
It should be noted that the fiber $\pi^{-1}(\infty)$ is smooth, hence
the group acts on it, and does not fix it. Hence the fix point set is
$\pi^{-1}(0) \cup \sigma_0$,
where $\sigma_0$ denotes the zero section. This will be used below.

\medskip
\noindent{$k=44$:} There is a Weierstra\ss\ equation
\beq
y^2=x^3+x+t^{11}
\eeq
and the automorphism is given by
$(x,y,t)\mapsto (\alpha_{44}^{22} x,\alpha_{44}^{11}y,\alpha_{44}^2t)$.
This elliptic fibration has a singular fiber of type \II\ over $t=\infty$ and
22 fibers of type I$_1$, over the roots of $t^{22}=-4/27$.

\medskip
\noindent{$k=42$:} Here we again have a description as an elliptic surface,
\beq
y^2=x^3+t^5\prod_1^7(t-\alpha_7^i)
\eeq
and the automorphism is given by
$(x,y,t) \mapsto (\alpha_{42}^2 x,\alpha_{42}^3 y,\alpha_{42}^{18}t)$.
It has a fiber of type $\II^*$ at $t=0$, and fibers of type $\II$ at all
seventh roots of unity. This example may also be described as a weighted
hypersurface:
\beq
\{x^2+y^3+z^7+w^{42}=0\}\subset \IP_{(1,6,14,21)}.
\eeq
Here the automorphism is given by
$(x,y,z,w) \mapsto (x,y,z,\alpha_{42}w)$.

The important point following from these remarks is
that we can now pose the problem the other way
around: given a gauge group of a heterotic string, we can
find a Calabi-Yau
with a K3 fibration such that the invariant lattice under the monodromy
is the lattice of the given gauge group. More precisely, suppose we can
find a K3 surface $K$, such that (i) it has $S_K$ given
by the lattice of the given
gauge group, and (ii) it has a non-trivial automorphism group $H_K$. Then
we can apply the construction above, and the result is a Calabi-Yau threefold,
fibered in K3 surfaces, such that the invariants under the monodromy are
exactly the lattice $S_K$.

Now let us try to find some interesting lattices which could play the role
of gauge groups for hererotic strings. Suppose, for example we are looking
for a type IIA string on a Calabi-Yau with gauge group $SO(8)$ and with
$(n_v,n_h)=(8,272)$. First we note that the following combination of
singular fibers would do the job: $1 \Inulls,\ 9 \II$; as mentioned above,
this would give a Picard lattice on a K3 $S_K=D_4 \oplus U$. We now
construct such an elliptic surface with an automorphism of order 18, as
follows. The Weierstra\ss\ equation will be
\beq
y^2=x^3+t^3\prod_1^9(t-\alpha_9^i),
\eeq
and an automorphism is given by
$(x,y,t)\mapsto (\alpha_{18}^2x,\alpha_{18}^3y, \alpha_{18}^2 t).$
{}From the general theory of elliptic surfaces, since
$g_3=t^3\prod_1^9(t-\alpha_9^i)$, which vanishes to order 3 at $t=0$ and
order 1 at $t=\alpha_{9}^i$, while $\Delta = g_2^3-27g_3^2=-27g_3^2$
vanishes to order 6 at $t=0$ and to order 2 at $t=\alpha_9^i$,
we find that there
are precisely the mentioned singular fibers, i.e., $1 \Inulls,\ 9 \II$.
We may do our construction above, using the curve $C_{18}$, and the result
is a Calabi-Yau with 36 singular fibers. Furthermore, the number $h^{1,1}$
is easy to find -- it is just one more than the corresponding value for the
K3 surface (as this automorphism is in $H_K$, the invariant lattice is just
$S_K$ which has rank 6), that is $h^{1,1}=7$. We now calculate the Euler
number of this fibration to find the other Hodge number. The singular
fibers are in this case also, cones over a reducible curve. On the elliptic
surface, this curve is given by the fibers over $0$
and the zero section. Note that this corresponds to the invariance of the
Picard lattice $S_K$, as $S_K$ is spanned by the classes: the fiber
$\pi^{-1}(0)$ and the pair (fiber,section), producing the lattice
$D_4\oplus U$. It is the curve on the elliptic surface
\[\unitlength .5cm
\begin{picture}(6,5)
\put(2,4.5){\line(0,-1){4.5}}
\put(1.5,4){\line(1,0){1}}
\put(1.5,3){\line(1,0){1}}
\put(1.5,2){\line(1,0){1}}
\put(1.5,1){\line(1,0){1}}
\put(1,.5){\line(1,0){4.5}}
\end{picture}
\]
and has Euler number $\chi(C)=7$. There are 36 singular fibers, and our
formula (\ref{fib-euler}) for the Euler number gives
$\chi(X)=-528$. Thus we find $(h^{1,1},h^{2,1})=(7,271)$, precisely as
needed.

Returning briefly to the example for $k=42$ recall the elliptic
fibration had $1 \II^*$ fiber and 7 of type $\II$, spanning a lattice
$E_8\oplus U$. These give the fixed curves under the action of
$\ZZ_{42}$, namely the fiber $\pi^{-1}(0)$ and the zero section, which is
the curve $C$ with Euler number 11 described in Section 3.
The Calabi-Yau constructed from this example has
84 singular fibers, and thus we recover our discussion of this example
in Section 3.

We may also apply the above construction to the two Kondo examples with
$k=44, 66$. In a sense, the situation of these examples is considerably
easier than with the case $k=42$, simply because there are not so
many rational curves. Indeed, as mentioned above, the fixed point set
under the automorprhism consists of the fiber $\pi^{-1}(0)$ and the zero
section for the example $k=66$, and for the example $k=44$ a similar
argument shows that the fixed curve consists of $\pi^{-1}(\infty)$ and
the zero section. In both cases, this is a singular fiber of type ${\bf II}$
and a smooth $\IP_1$, so has Euler number 3. Applying our formula
(\ref{fib-euler}) for $\chi$ of the Calabi-Yau, we get
\bea
k=44 &:& \quad \chi(X)=(2-88)\cdot 24 +88\cdot 4 = -1712, \nn \\
k=66 &:& \quad \chi(X)=(2-132)\cdot 24 + 132\cdot 4 = -2592.
\eea
These two examples hence have an Euler number far smaller than all examples
known up till now, and also the highest number of singular fibers of any
Calabi-Yau threefold with K3 fibration known to date.

\vskip .2truein
\noindent
{\bf Acknowledgement}

\noindent The first author is indebted to Igor Dolgachev for discussions on
the examples of Kondo. The second author
is grateful to Per Berglund, Shyamoli Chaudhuri, Jens Erler, David Lowe
and Andy Strominger for discussions.
This work was supported in part  by NSF grant PHY--94--07194.

\end{document}